# Sphaleron transition rate at high temperature in the 1+1 D abelian Higgs model

J. Smit and W.H. Tang *

Institute of Theoretical Physics, Valckenierstraat 65, 1018 XE Amsterdam, the Netherlands

New results for the rate are presented using the canonical ensemble in the classical approximation on a spatial lattice. We find that the rate at high temperatures is proportional to $T^2$, and strongly dependent on the lattice spacing $a$. We conclude that a better effective action is needed for the classical approximation.

## 1. Introduction

The rate $\Gamma$ of baryon number violation in the Standard Model is difficult to calculate at finite temperature $T$. Therefore, the classical approximation was proposed which allows for making a numerical estimate. It was tested in the 1+1 dimensional abelian Higgs model [1–4], for which a semiclassical calculation of the rate is available,

$$F(\frac{T}{m_\phi}, \xi, v) \equiv \frac{\Gamma}{m_\phi^2 L} = \mathrm{f}(\xi)\sqrt{\frac{E_s}{T}} \exp\left[-\frac{E_s}{T}\right]. \quad (1)$$

Here $L$ is the one dimensional volume, $m_\phi$ is the Higgs mass, $v^{-2} = 2\lambda/m_\phi^2$ is the effective dimensionless coupling, $\xi = g^2/\lambda$, $g$ is the gauge coupling, $\lambda$ is the quartic Higgs self coupling, $E_s = (2/3)v^2 m_\phi$ is the classical sphaleron energy, and $f(\xi)$ is calculated in [5]. This formula is expected to be valid at temperatures $m_\phi \ll T \ll E_s$ which is confirmed by numerical simulations [4,3].

At high temperatures the rate is still unknown even for this rather simple model. From dimensional arguments one expects that the rate behaves as $T^2$,

$$F = \kappa(\xi, v) \frac{(v^{-2}T)^2}{m_\phi^2} \quad \text{for } T \gg E_s. \quad (2)$$

## 2. Classical approximation

At high temperature one expects high occupation numbers for quanta with energies much less than $T$ and for their contribution a classical approximation seems reasonable. Suppose we derive an effective action in which the spatial momenta are limited to $|p| < \Lambda$. Then the effective energies will be restricted by $\Lambda$ which suggests a classical approximation based on this effective action for $\Lambda \ll T$. Such an effective action $\bar{S}$ might be defined (in a generic field theory with generic fields $\varphi$) as follows,

$$\exp\left[\bar{S}(\bar{\varphi}, \Lambda)\right] = \int \mathrm{D}\varphi \, B_\Lambda(\bar{\varphi}, \varphi) \exp\left[S(\varphi)\right]. \quad (3)$$

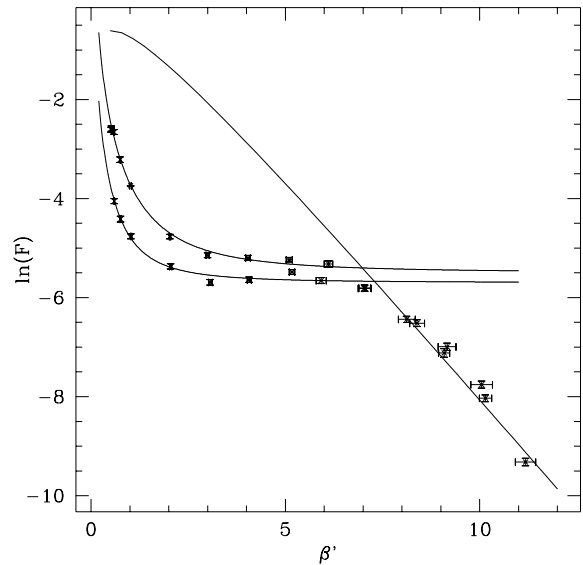

Figure 1. Results for $\ln F = \ln(\Gamma/m_\phi^2 L)$ for $\xi = 4$. The diagonal line represents the analytic formula (1). The other two lines are fits to the form $c_0 + c_2 T'^2$. The upper data are for $a' = 0.32$, the lower data for $a' = 0.16$.

Here $B_\Lambda$ is a blocking function such that $\int \mathrm{D}\bar{\varphi} \, B_\Lambda(\bar{\varphi}, \varphi) = 1$, e.g.

---
*Speaker at the conference

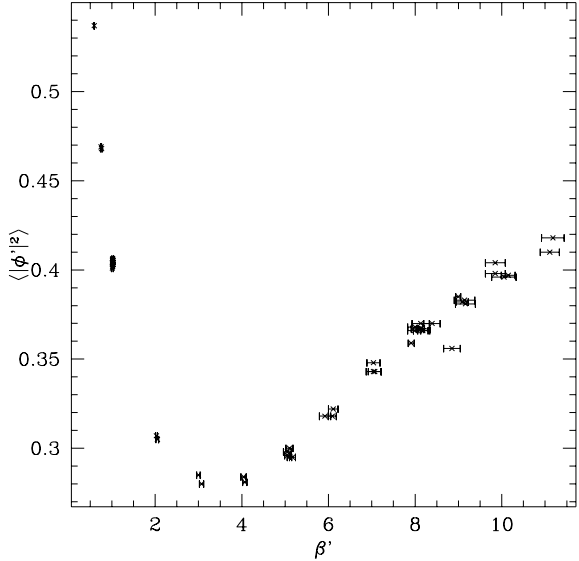

Figure 2. $\langle|\phi'|^2\rangle$ versus inverse temperature $\beta'$.

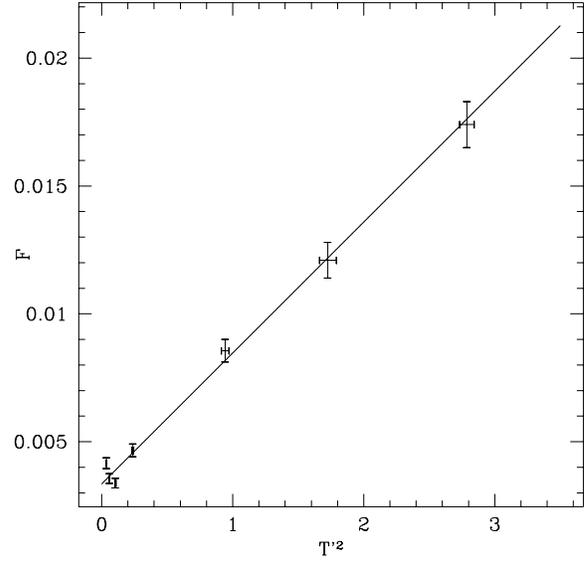

Figure 3. $F$ versus $T'^2$ for $a' = 0.16$. The line represents the fit $F = 0.00335 + 0.00512\, T'^2$.

$$B_\Lambda(\bar\varphi, \varphi) = \prod_{(x,t)} \delta[\bar\varphi(x,t) - \int_{|p|<\Lambda} dp\, \tilde\varphi(p,t) e^{ipx}]. \quad (4)$$

In this spirit we interpret the effective action for the abelian Higgs model on a lattice with lattice spacing $a$,

$$\begin{aligned}\bar S &= \int dt\, a \sum_n \\ &\quad \left[ \frac{1}{4g^2}\left(\frac{A_{0\,n+1} - A_{0\,n}}{a} - \partial_t A_{1\,n}\right)^2 \right. \\ &\quad - \frac{1}{a^2}|\exp(-iaA_{1\,n})\phi_{n+1} - \phi_n|^2 \\ &\quad \left. + |(\partial_t - iA_{0\,n})\phi_n|^2 - \lambda\left(|\phi_n|^2 - \frac{v^2}{2}\right) \right], \end{aligned} \quad (5)$$

as an approximation to such an effective action (we omitted the 'bar' on the fields, $\Lambda = \pi/a$). For convenience we make a scale transformation to dimensionless variables [2,3], $a = a'/v\sqrt\lambda$, $t = t'/v\sqrt\lambda$, $\phi = v\phi'$, $A_\mu = v\sqrt\lambda A'_\mu$, $m_\phi = \sqrt\lambda v m'_\phi$, $H = \sqrt\lambda v^3 H'$, $T = \sqrt\lambda v^3 T'$, $\beta' = 1/T'$, where $H$ is the hamiltonian.

### 3. Results

Fig. 1 gives an overview of the rate obtained in our simulation, which was carried out in the same way as in [3]. The classical approximation is consistent with the semiclassical calculation in the semiclassical region [3,4]. We see also an intermediate region which is almost flat. This region appears to correspond to the minimum of $\langle|\phi'|^2\rangle$ in fig. 2. Figs. 3 and 4 show more detailed data at high temperature. We see here a $T'^2$ behavior which starts already just beyond the minimum of $\langle|\phi'|^2\rangle$. According to fig 1, the rate at high temperature depends on the lattice spacing $a'$. With a gauge invariant lattice action for bosonic fields one expects the lattice spacing dependence to be like a series in $a'^2$. This is confirmed in figs. 5 and 6 for two values of $T'$. A linear $a'^2$ approximation appears to hold for $a'^2 \lesssim 0.1$. We therefore fit the data in figs. 3 and 4 to the following ansatz,

$$\begin{aligned}F &= c_{00} + c_{02}\, a'^2 + (c_{20} + c_{22}\, a'^2) T'^2 \\ &= c_{00} + c_{02}\frac{a^2 m_\phi^2}{2} \\ &\quad + 2\left(c_{20} + c_{22}\frac{a^2 m_\phi^2}{2}\right)\frac{(v^{-2}T)^2}{m_\phi^2},\end{aligned} \quad (6)$$

and find

$$c_{00} = (3.1 \pm 0.1)10^{-3},\ c_{02} = (9.6 \pm 1.1)10^{-3}, \\ c_{22} = 0.20 \pm 0.02, \qquad c_{20} = (-7 \pm 25)10^{-5}. \quad (7)$$

From this we see that the $v^{-4}\, T^2$ behavior in eq. 2 is confirmed (contradicting [6]), with

$$\kappa(\xi, v) \approx 0.20\, a^2\, m_\phi^2, \quad \text{at } \xi = 4. \tag{8}$$

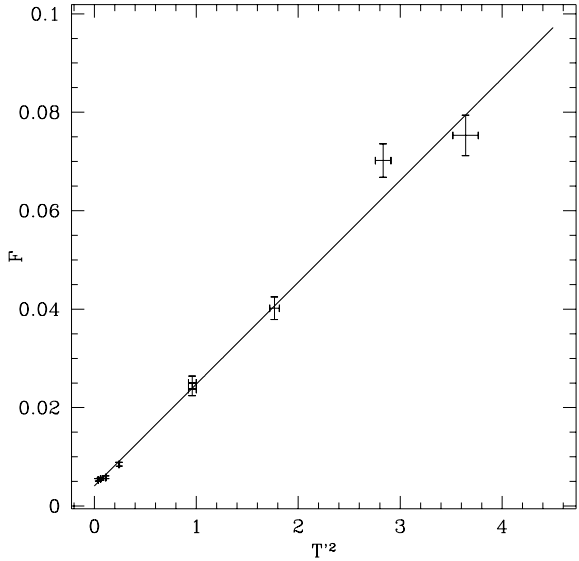

Figure 4. $F$ versus $T'^2$ for $a' = 0.32$ and the fit $F = 0.00409 + 0.0207\, T'^2$.

The coefficient of $T^2$ depends strongly on the lattice scale. Therefore, it is absolutely necessary to understand the $a$ dependence of the coefficients in the effective action before we can say to have calculated $\kappa$.

**Acknowledgements:** We would like to thank A.I. Bochkarev and A. Krasnitz for useful discussions. This work is financially supported by the Stichting voor Fundamenteel Onderzoek der Materie (FOM) and the Stichting Nationale Computer Faciliteiten (NCF).

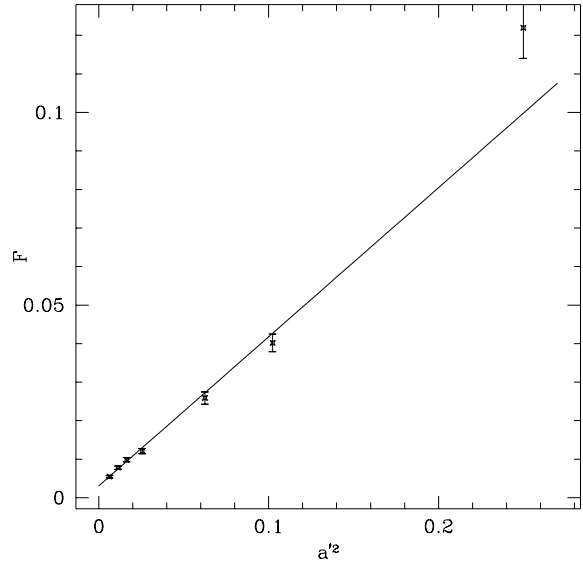

Figure 5. $F$ versus $a'^2$ for $T' = 4/3$. The line represents the fit $F = 0.00307 + 0.387 a'^2$.

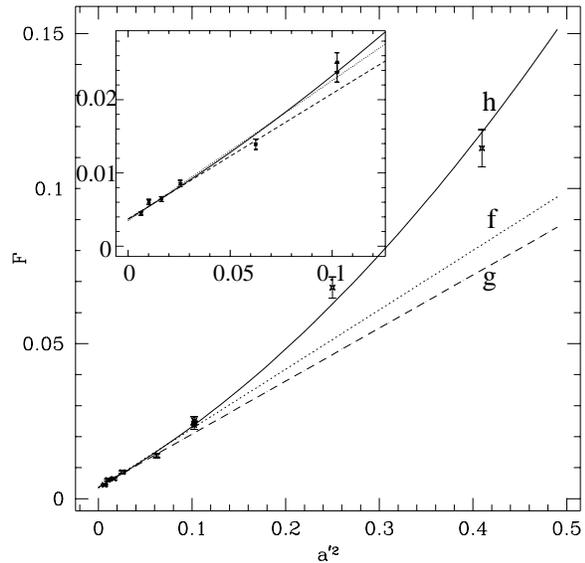

Figure 6. $F$ versus $a'^2$ for $T' = 1$, and the fits $f = 0.00347 + 0.192\, a'^2$ to the seven lowest data points, $g = 0.00373 + 0.171 a'^2$ to the five lowest data points and $h = 0.00367 + 0.169\, a'^2 + 0.270\, a'^4$ to all data. The insert zooms in on the region $a'^2 \lesssim 0.1$.